\documentclass[preprint,12pt]{elsarticle}
\usepackage{lineno}
\usepackage{graphicx}
\usepackage{bm}
\usepackage{amsmath,upgreek}
\usepackage{amssymb}
\usepackage{hyperref}
\hypersetup{colorlinks=true,allcolors=blue}
\modulolinenumbers[5]

\journal{CNSNS}

\begin{document}

\begin{frontmatter}

\title{Exact stationary solutions of the Kolmogorov-Feller equation in a bounded domain}
\author{S.I.\ Denisov}\corref{corr}
\ead{denisov@sumdu.edu.ua}
\cortext[corr]{Corresponding author}
\author{Yu.S.\ Bystrik}
\address{Sumy State University, 2 Rimsky-Korsakov Street, UA-40007
Sumy, Ukraine}

\begin{abstract}
We present the first detailed analysis of the statistical properties of jump processes bounded by a saturation function and driven by Poisson white noise, being a random sequence of delta pulses. The Kolmogorov-Feller equation for the probability density function (PDF) of such processes is derived and its stationary solutions are found analytically in the case of the symmetric uniform distribution of pulse sizes. Surprisingly, these solutions can exhibit very complex behavior arising from both the boundedness of pulses and processes. We show that all features of the stationary PDF (number of branches, their form, extreme values probability, etc.) are completely determined by the ratio of the saturation function width to the half-width of the pulse-size distribution. We verify all theoretical results by direct numerical simulations.
\end{abstract}


\end{frontmatter}

\linenumbers

\section{Introduction}
\label{Intr}

The Langevin equation (i.e., a sto\-chastic ordinary differential equation) is widely used for studying stochastic systems in physics, chemistry, engineering and other areas \textcolor{blue} {\cite{CoKaWa2004}}. In the simplest case when the random force noise is Gaussian and white the dynamics of the system is Markovian and its probability density function (PDF) satisfies the Fokker-Planck equation \textcolor{blue}{\cite{Risk1989, HoLe2006, Gard2009}}. One of the advantages of this approach is that the Fokker-Planck equation can often be solved analytically, especially in the stationary regime.

If Gaussian noise is colored, then the system dynamics becomes non-Markovian and the corresponding PDF obeys the integro-differential master equation, which under certain conditions can be reduced to the differential one by the Kramers-Moyal expansion \textcolor{blue}{\cite{CoKaWa2004, Risk1989}}. Since, in general, this differential equation is of infinite order, several approximation schemes for its simplification were proposed \textcolor{blue} {\cite{SaMi1989, LiWe1990, HaJu1995}} (for a recent theoretical and numerical analysis see, e.g., Refs.\ \textcolor{blue} {\cite{WaTaTa_PRL2013, MaGrTa_JCP2018}} and references therein). Note also that in some very special cases when the Langevin equation is solved analytically the PDFs can be determined straightforwardly \textcolor{blue} {\cite{HoLe2006, DeHo_PRE2002a, DeHo_PRE2002b, Vitr_PA2006}}.

The Langevin equation driven by Poisson white noise (sometimes called a train of delta pulses), which is a particular case of non-Gaussian white noises, plays an important role in describing the jump processes and phenomena induced by this noise in different systems (see, e.g., Refs.\ \textcolor{blue} {\cite{Han_ZPhys1980, HePeRo_PRE1987, LuBaHa_EPL1995, Grig2002, Gitt2005}}). More recent studies include noise-induced transport \textcolor{blue} {\cite{BaSo_PRE2013, SHL_PRE2014}}, stochastic resonance \textcolor{blue} {\cite{HXSJ_ND2017}}, vibro-impact response \textcolor{blue} {\cite{Zhu_ND2015, YXHH_MPE2018}} and ecosystem dynamics \textcolor{blue}{\cite{PZ_AMS2014, JXL_Entr2018}}, to name only a few. The determination of the corresponding PDF is a much more difficult problem than for Gaussian white noise, because the master equation is integro-differential. Note in this connection that even for the first-order Langevin equation the master equation reduces to the integro-differential Kolmogorov-Feller equation, whose exact stationary solutions are known only in a few cases \textcolor{blue} {\cite{Vast_IJNLM1995, Prop_IJNLM2003, DeHoHa_EPJB2009, RuDuGu_DM2016, DuRuGu_PRE2016}}.

Often the bounded processes more adequately describe the stochastic behavior of real systems than the unbounded ones \textcolor{blue}{\cite{d'Onofr2013}}. But the bounded jump processes driven by Poisson white noise, which could be used, for example, to model the destruction phenomena, have not been studied in depth. As far as we know, our recent paper \textcolor{blue} {\cite{DeBy_PhysA2018}} is the only one devoted to the analytical study of the statistical properties of such processes. It has been shown, in particular, that the jump character and boundedness of these processes are responsible for the nonzero probability of their extremal values and nonuniformity of their distribution inside a bounded domain.

In this work, we generalize the difference Langevin equation describing bounded jump processes driven by Poisson white noise, derive the corresponding Kolmogorov-Feller equation and solve it analytically in the stationary state for the case of uniform distribution of pulse sizes. The paper is organized as follows. In \textcolor{blue}{Section \ref{Model}}, using the saturation function, we introduce the difference Langevin equation driven by Poisson white noise, whose solutions are bounded. The Kolmogorov-Feller equation that corresponds to this Langevin equation is derived in \textcolor{blue}{Section \ref{KFeq}}. In the same section, we cast the stationary solution of the Kolmogorov-Feller equation as a sum of singular terms defining the probability of the extremal values of the bounded process and a regular part representing the non-normalized PDF of this process inside a bounded domain. In \textcolor{blue}{Section \ref{ExSol}}, which is the main section of the paper, we solve analytically the integral equation for the non-normalized PDF and calculate the extreme values probability in the case of uniform distribution of pulse sizes. Here, we show that the ratio of the saturation function width to the half-width of the pulse-size distribution is the only parameter that determines all features of the non-normalized PDF, including its explicit form and complexity. Finally, our main findings are summarized in  \textcolor{blue}{Section \ref{Concl}}.

\section{Model for bounded stochastic processes}
\label{Model}

A variety of continuous-time processes in physics, biology, economics and other areas can be described by the first-order Langevin equation
\begin{equation}
    \frac{d}{dt}X_{t} = F(X_{t}) + \xi(t),
    \label{Lang}
\end{equation}
which, for convenience, is often written in difference form
\begin{equation}
    X_{t + \tau} = X_{t} + F(X_{t})\tau +
    \Delta_{\tau}.
    \label{Lang_dif}
\end{equation}
Here, $X_{t}$ ($t \geq 0$) is a random process, $F(x)$ is a giving deterministic function, $\xi(t)$ is a stationary white noise, $\tau$ is an infinitesimal time interval, and $\Delta_{\tau}$ is a random variable defined as
\begin{equation}
    \Delta_{\tau} = \int_{t}^{t+\tau}
    \xi(t')\,dt' = \int_{0}^{\tau}\xi(t')\,dt'.
    \label{def_Delta}
\end{equation}

The realizations of $X_{t}$ can be either continuous (as in the case of Gaussian white noise) or discontinuous (as in the cases, e.g., of L\'{e}vy and Poisson white noises). These realizations are, in general, unbounded, i.e., the probability that $|X_{t}|$ exceeds a given level is nonzero. In order to extend the Langevin approach to the description of random processes in bounded domains, we introduce instead of \textcolor{blue}{Eq.\ (\ref{Lang_dif})} a more general difference Langevin equation
\begin{equation}
    X_{t + \tau} = S(X_{t} + F(X_{t})\tau +
    \Delta_{\tau}),
    \label{eq_X}
\end{equation}
where
\begin{equation}
    S(x) = \left\{\! \begin{array}{ll}
    x, & |x| \leq l,
    \\ [6pt]
    \mathrm{sgn}(x)\,l, & |x| > l
    \end{array}
    \right.
    \label{S(x)}
\end{equation}
is the saturation function, $2l$ is its width (domain size), and $\mathrm{sgn} (x)$ is the signum function. According to \textcolor{blue}{Eq.\ (\ref{eq_X})} and definition \textcolor{blue}{(\ref{S(x)})}, a nonlinear random process $X_{t}$ is bounded, i.e., if $X_{0} \in [-l, l]$, then $X_{t}$ evolves in such a way that $|X_{t}| \leq l$ for all $t \geq 0$. Note, this equation reduces to \textcolor{blue}{Eq.\ (\ref{Lang_dif})} when $l \to \infty$.

Although in \textcolor{blue}{Eq.\ (\ref{eq_X})} any noise can be used, next we explore Poisson white noise only, which is defined as a sequence of delta pulses (see, e.g., Ref.\ \textcolor{blue} {\cite{Grig2002}} and references therein):
\begin{equation}
    \xi(t) = \sum_{i=1}^{n(t)}z_{i}
    \delta(t-t_{i}).
    \label{def_xi}
\end{equation}
Here, $n(t)$ denotes the Poisson counting process, which is characterized by the probability $Q_{n}(t) = (\lambda t)^{n} e^{-\lambda t}/ n!$ that $n \geq 0$ events occur at random times $t_{i}$ within a given time interval $(0, t]$, $\lambda$ is the rate parameter, $\delta(\cdot)$ is the Dirac $\delta$ function, and $z_{i}$ are independent random variables distributed with the same probability density $q(z)$ [$z \in (-\infty, \infty)$]. It is also assumed that this probability density is symmetric, $q(-z) = q(z)$, and $\xi(t) =0$ if $n(t) =0$. From \textcolor{blue} {(\ref{def_Delta})} and \textcolor{blue} {(\ref{def_xi})} it follows that in the case of Poisson white noise the random variable $\Delta_{\tau}$ is the compound Poisson process \textcolor{blue}{\cite{Grig2002}}, i.e.,
\begin{equation}
    \Delta_{\tau} =
    \left\{\! \begin{array}{ll}
    0, & n(\tau)=0,
    \\
    \sum\nolimits_{i=1}
    ^{n(\tau)}\!z_{i},
    & n(\tau) \geq 1.
    \end{array}
    \right.
    \label{Delta}
\end{equation}
Since $\tau \to 0$, the probability density $p_{\tau}(z)$ that $\Delta_{\tau} = z$ is written in the linear approximation in $\tau$ as \textcolor{blue}{\cite{DeHoHa_EPJB2009}}
\begin{equation}
    p_{\tau}(z) = (1 - \lambda \tau) \delta(z)
     + \lambda \tau q(z).
    \label{p2}
\end{equation}

\section{Kolmogorov-Feller equation}
\label{KFeq}

\subsection{Time-depended case}

Our next aim is to derive the Kolmogorov-Feller equation for the normalized time-depended PDF $P_{t}(x)$ of the bounded process $X_{t}$ governed by \textcolor{blue}{Eq.\ (\ref{eq_X})}. Using the definition $P_{t}(x) = \langle \delta(x - X_{t}) \rangle$, where $x \in [-l, l]$, the angular brackets denote averaging over all realizations of $X_{t}$, and two-step averaging procedure for $\langle \delta(x - X_{t+\tau}) \rangle$ \textcolor{blue} {\cite{DeViHo_PRE2003}}, we can write
\begin{align}
    P_{t+\tau}(x) &=  \langle
    \delta [x - S(X_{t} + F(X_{t})\tau + \Delta_{\tau})]
    \rangle
    \nonumber \\[3pt]
    &= \int_{-l}^{l}\! P_{t}(x')\,
    \Bigg(\! \int_{-\infty}^{\infty}\! p_{\tau}(z)
    \delta [x - S(x' + F(x')\tau + z)]\,dz
    \Bigg)\,dx'.
    \label{P1}
\end{align}
Taking also into account the representation
\begin{equation}
    P_{t}(x)  = \int_{-l}^{l}\! P_{t}(x')\,
    \Bigg(\! \int_{-\infty}^{\infty}\! p_{\tau}(z)
    \delta (x - x')\,dz \Bigg)\,dx'
    \label{P2}
\end{equation}
(it holds due to the normalization condition $\int_{-\infty}^ {\infty} p_{\tau}(z)\,dz = 1$ and shifting property of the $\delta$ function) and the definition $(\partial/ \partial t)P_{t}(x) = \lim_{\tau \to 0} [P_{t+\tau}(x) - P_{t}(x)]/ \tau$, from \textcolor{blue}{(\ref{P1})} and \textcolor{blue} {(\ref{P2})} one obtains
\begin{equation}
    \frac{\partial}{\partial t}P_{t}(x) =
    \int_{-l}^{l}K(x,x')P_{t}(x')\,dx',
    \label{eq_P}
\end{equation}
where
\begin{equation}
    K(x,x') = \lim_{\tau \to 0} \frac{1}
    {\tau}\! \int_{-\infty}^{\infty}p_{\tau}
    (z) \{ \delta [x - S(x' + F(x')\tau + z)]
    - \delta (x-x')\}\,dz
    \label{K1}
\end{equation}
($x,x' \in [-l,l]$) is the kernel of the master equation \textcolor{blue}{(\ref{eq_P})}.

In order to derive the Kolmogorov-Feller equation associated with \textcolor{blue}{Eq.\ (\ref{eq_X})} at $\tau \to 0$, we first substitute the probability density \textcolor{blue} {(\ref{p2})} into \textcolor{blue} {(\ref{K1})}. After integration over $z$ one gets
\begin{align}
    K(x,x') = &\lim_{\tau \to 0} \frac{1}
    {\tau} \bigg\{ (1 - \lambda \tau)\delta
    [x - S(x' + F(x')\tau)] - \delta (x - x')
    \nonumber \\[3pt]
    &+ \lambda \tau \int_{-\infty}^{\infty}
    q(z) \delta [x - S(x' + F(x')\tau + z)]\,dz
    \bigg\}.
    \label{K1_1}
\end{align}
Then, replacing $S(x' + F(x')\tau + z)$ by $S(x' + z)$ (this is possible because only terms of the order of $\tau$ in braces contribute to the limit) and taking into account that $S(x') = x'$ and, in the linear approximation,
\begin{equation}
    \delta[x - S(x' + F(x')\tau)] = \delta
    (x - x') - \tau \frac{\partial} {\partial x}
    \delta(x - x') F(x'),
    \label{expr}
\end{equation}
the kernel \textcolor{blue} {(\ref{K1_1})} can be rewritten in the form
\begin{equation}
    K(x,x') = -\frac{\partial} {\partial x}
    \delta(x - x')F(x') - \lambda \delta
    (x - x') + \lambda \int_{-\infty}^{\infty}
    \!q(z) \delta [x - S(x' + z)]\,dz.
    \label{K1_2}
\end{equation}

Finally, using in \textcolor{blue} {(\ref{K1_2})} the representation
\begin{align}
    \int_{-\infty}^{\infty}q(z) \delta
    [x - S(x' + z)]\,dz  =
    & \, \delta (x + l) \int_{-\infty}^
    {-l-x'}q(z) dz + \delta (x - l)
    \int_{l-x'}^{\infty}q(z)\,dz
    \nonumber \\[3pt]
    & + \int_{-l - x'}^{l - x'}
    q(z) \delta (x - x' - z)\,dz,
    \label{int}
\end{align}
which directly follows from the definition \textcolor{blue} {(\ref{S(x)})} of the saturation function, the integral formula $\int_{-l - x'}^{l - x'} q(z) \delta (x - x' - z)\,dz = q(x - x')$, and the exceedance probability defined as
\begin{equation}
    R(z) = \int_{z}^{\infty} q(z')\,dz'
    \label{def_R}
\end{equation}
[$R(-\infty)=1$, $R(0)=1/2$, $R(\infty)=0$], we obtain
\begin{align}
    K(x,x') = &- \frac{\partial}{\partial x}
    \delta(x-x') F(x') - \lambda \delta(x-x')
    + \lambda \delta(x-l) R(l-x')
    \nonumber \\[3pt]
    & + \lambda \delta(x+l)R(l+x') +
    \lambda q(x-x').
    \label{K2}
\end{align}

Now, substituting this kernel into \textcolor{blue}{Eq.\ (\ref{eq_P})}, we get the Kolmogorov-Feller equation
\begin{align}
    \frac{1}{\lambda}\frac{\partial}
    {\partial t} P_{t}(x) &+  \frac{1}
    {\lambda}\frac{\partial}{\partial x}
    F(x) P_{t}(x) + P_{t}(x) = \delta(x-l)
    \!\int_{-l}^{l} R(l-x') P_{t}(x')\,dx'
    \nonumber \\[3pt]
    &+ \delta(x+l)\! \int_{-l}^{l} R(l+x')
    P_{t}(x')\,dx' + \int_{-l}^{l} q(x-x')
    P_{t}(x')\,dx',
    \label{KF}
\end{align}
which corresponds to the difference Langevin equation \textcolor{blue} {(\ref{eq_X})} with $\tau \to 0$ (note, the Kolmogorov-Feller equation for $F(x)=0$ has been derived in Ref.\ \textcolor{blue} {\cite{DeBy_PhysA2018}}). As usual, \textcolor{blue}{Eq.\ (\ref{KF})} should be supplemented by the normalization, $\int_{-l}^{l} P_{t}(x)\,dx = 1$, and initial, $P_{0}(x) = \delta(x-X_{0})$, conditions. It should also be emphasized that, according to \textcolor{blue} {\cite{DeBy_PhysA2018}}, any boundary conditions at $x=\pm l$ are not needed to solve this equation.

\subsection{Stationary PDF and its representation}

Our future efforts will be focused only on the stationary PDF $P_{\mathrm{st}}(x) = \lim_{t \to \infty} P_{t}(x)$ at $F(x)=0$. Since by assumption $q(-z) = q(z)$, in this case the stationary PDF is symmetric, $P_{\mathrm{st}} (-x) = P_{\mathrm{st}} (x)$, and, as it follows from \textcolor{blue}{Eq.\ (\ref{KF})}, satisfies the integral equation
\begin{equation}
    P_{\mathrm{st}}(x) = [\delta(x-l) +
    \delta(x+l)]\! \int_{-l}^{l} R(l-x')
    P_{\mathrm{st}}(x')\,dx' + \!\int_{-l}^{l}
    q(x-x')P_{\mathrm{st}}(x')\,dx'.
    \label{KFst}
\end{equation}

According to \textcolor{blue}{\cite{DeBy_PhysA2018}}, the general solution of \textcolor{blue}{Eq.\ (\ref{KFst})} can be represented in the form
\begin{equation}
    P_{\mathrm{st}}(x) = a[\delta(x-l) +
    \delta(x+l)] + f(x),
    \label{Pst}
\end{equation}
where $a$ is the probability that $X_{t}$ in the stationary state equals $l$ (or $-l$), and the non-normalized probability density $f(x)$ is symmetric, $f(-x) = f(x)$, and is governed by the integral equation
\begin{equation}
    f(x) = a[q(x-l) + q(x+l)] + \int_{-l}^{l}
    q(x-x')f(x')\,dx'.
    \label{f_eq}
\end{equation}
Using \textcolor{blue} {(\ref{Pst})} and the normalization condition $\int_{-l}^{l} P_{\mathrm{st}}(x)\,dx = 1$, the probability $a$ of the extremal values of the process $X_{t}$ in the stationary state can be expressed through the non-normalized PDF $f(x)$ as follows:
\begin{equation}
    a = \frac{1}{2} - \int_{0}^{l} f(x)\,dx.
    \label{def_a}
\end{equation}

\section{Exact solutions for uniform jumps}
\label{ExSol}

\subsection{Basic equations}
\label{BasEq}

In order to solve \textcolor{blue}{Eq.\ (\ref{f_eq})} analytically, we restrict ourselves to the case when the jump magnitudes $z_{i}$ are uniformly distributed on the interval $[-c, c]$ ($c>0$ is the half-width of this distribution). In other words, we assume that the probability density $q(z)$ is given by
\begin{equation}
    q(z) = \left\{\! \begin{array}{ll}
    1/2c, & |z| \leq c,
    \\
    0, & |z| > c.
    \end{array}
    \right.
    \label{q(z)}
\end{equation}
Depending on the value of $c$, \textcolor{blue}{Eq.\ (\ref{f_eq})} can be rewritten in three different forms. First, if $c>2l$, then
\begin{equation}
    q(x-l) = q(x+l) = q(x-x') = \frac{1}{2c}
    \label{q1}
\end{equation}
for all $x,x' \in [-l, l]$, and \textcolor{blue}{Eq.\ (\ref{f_eq})} reduces to
\begin{equation}
    f(x) = \frac{a}{c} + \frac{1}{c}
    \int_{0}^{l}f(x')\,dx'.
    \label{f_eq1}
\end{equation}

Second, if $c \in (l, 2l)$, then
\begin{equation}
    q(x-l) = \left\{\! \begin{array}{ll}
    0, & x \in [-l, l-c),
    \\
    1/2c, & x \in [l-c, l],
    \end{array}
    \right.
    \quad
    q(x+l) = \left\{\! \begin{array}{ll}
    1/2c, & x \in [-l, c-l],
    \\
    0, & x \in (c-l, l]
    \end{array} \label{q2}
    \right.
\end{equation}
and
\begin{equation}
    \int_{-l}^{l} q(x-x')f(x')\,dx' =
    \frac{1}{2c} \times
    \left\{\! \begin{array}{ll}
    \int_{-l}^{x+c}f(x')\,dx',
    & x \in [-l, l-c],
    \\ [3pt]
    \int_{-l}^{l}f(x')\,dx',
    & x \in [l-c, c-l],
    \\ [3pt]
    \int_{x-c}^{l}f(x')\,dx',
    & x \in [c-l, l].
    \end{array}
    \right.
    \label{Int_q2}
\end{equation}
Using these results, from \textcolor{blue}{Eq.\ (\ref{f_eq})} one obtains the following integral equations:
\begin{subequations}
    \label{f_eq2}
    \begin{equation}
    \label{f_eq2a}
    f(x) = \frac{a}{2c} + \frac{1}{2c}
    \int_{-l}^{x+c}f(x')\,dx'
    \end{equation}
\end{subequations}
for $x \in [-l, l-c)$,
\begin{displaymath} \tag{29b}
    f(x) = \frac{a}{c} + \frac{1}{2c}
    \int_{-l}^{l}f(x')\,dx'
    \label{f_eq2b}
\end{displaymath}
for $x \in (l-c, c-l)$, and
\begin{displaymath} \tag{29c}
    f(x) = \frac{a}{2c} + \frac{1}{2c}
    \int_{x-c}^{l}f(x')\,dx'
    \label{f_eq2c}
\end{displaymath}
for $x \in (c-l, l]$.

And third, if $c \in (0, l)$, then the probability densities $q(x-l)$ and $q(x-l)$ are given by the same formulas \textcolor{blue} {(\ref{q2})}, and
\begin{equation}
    \int_{-l}^{l} q(x-x')f(x')\,dx' =
    \frac{1}{2c} \times
    \left\{\! \begin{array}{ll}
    \int_{-l}^{x+c}f(x')\,dx',
    & x \in [-l, c-l],
    \\ [3pt]
    \int_{x-c}^{x+c}f(x')\,dx',
    & x \in [c-l, l-c],
    \\ [3pt]
    \int_{x-c}^{l}f(x')\,dx',
    & x \in [l-c, l].
    \end{array}
    \right.
    \label{Int_q3}
\end{equation}
Hence, in this case \textcolor{blue}{Eq.\ (\ref{f_eq})} yields
\begin{subequations}
    \label{f_eq3}
    \begin{equation}
    \label{f_eq3a}
    f(x) = \frac{a}{2c} + \frac{1}{2c}
    \int_{-l}^{x+c}f(x')\,dx'
    \end{equation}
\end{subequations}
for $x \in [-l, c-l)$,
\begin{displaymath} \tag{31b}
    f(x) = \frac{1}{2c}
    \int_{x-c}^{x+c}f(x')\,dx'
    \label{f_eq3b}
\end{displaymath}
for $x \in (c-l, l-c)$, and
\begin{displaymath} \tag{31c}
    f(x) = \frac{a}{2c} + \frac{1}{2c}
    \int_{x-c}^{l}f(x')\,dx'
    \label{f_eq3c}
\end{displaymath}
for $x \in (l-c, l]$.

A remarkable advantage of \textcolor{blue}{Eqs.\ (\ref{f_eq1})}, \textcolor{blue} {(\ref{f_eq2})} and \textcolor{blue} {(\ref{f_eq3})} is that they can be solved analytically and, what is especially important, the choice of $q(z)$ in the form \textcolor{blue} {(\ref{q(z)})} permits us to characterize the complexity of the function $f(x)$ by a single ratio parameter $\sigma = 2l/c$. In particular, it will be demonstrated that, if $\sigma \in (n-1, n)$ with $n= \overline{1, \infty}$, then $f(x)$ is a piecewise continuous function, which, in general, consists of $2n-1$ branches. These branches are separated from each other by $2(n-1)$ points $\pm x_{k}$, where $k = \overline{1, n-1}$ ($n\geq 2$) and $x_{k} = |2k/\sigma -1|l$, at which the function $f(x)$ can be either continuous or discontinuous (with jump discontinuity). The change of the number of branches occurs at the critical values $\sigma_{\mathrm{ cr}} = n-1$ of the ratio parameter $\sigma$. Next, we determine the function $f(x)$ for $n=1,2,3$ and $n \to \infty$, calculate the probability $a$, and compare analytical results with those obtained by numerical simulations of \textcolor{blue}{Eq.\ (\ref{eq_X})}.

\subsection{Solution at \texorpdfstring{$\sigma \in (0,1)$}{Lg}}

The condition $n=1$ [i.e., $\sigma \in (0,1)$] means that $c>2l$ and hence the function $f(x)$ obeys \textcolor{blue}{Eq.\ (\ref{f_eq1})}, according to which $f(x) = f = \mathrm{const}$. The substitution of $f(x) = f$ into \textcolor{blue}{Eq.\ (\ref{f_eq1})} and condition \textcolor{blue}{(\ref{def_a})} yields a set of equations $f = a/c + fl/c$ and $a = 1/2 -fl$. Solving it with respect to $f$ and $a$ and introducing the reduced non-normalized probability density $\tilde{f}( \tilde{x})$ as $\tilde{f} (\tilde{x}) = f(l\tilde{x})l$\, ($\tilde{x} = x/l$) and $\tilde{f}$ as $\tilde{f} = fl$, we obtain
\begin{equation}
    \tilde{f} = \frac{\sigma}{4},
    \quad
    a= \frac{1}{2} - \frac{\sigma}{4}.
    \label{f1,a1}
\end{equation}
From this, using a general representation
\begin{equation}
    \tilde{P}_{\mathrm{st}}(\tilde{x}) =
    a[\delta(\tilde{x}-1) + \delta(
    \tilde{x}+1)] + \tilde{f} (\tilde{x})
    \label{red_Pst}
\end{equation}
of the reduced PDF $\tilde{P}_{\mathrm{st}} (\tilde{x}) = P_{\mathrm{st}} (l\tilde{x})l$, one gets
\begin{equation}
    \tilde{P}_{\mathrm{st}}(\tilde{x}) =
    \bigg( \frac{1}{2} - \frac{\sigma}{4}
    \bigg)\, [\delta(\tilde{x}-1) + \delta(
    \tilde{x}+1)] + \frac{\sigma}{4}.
    \label{Pst1}
\end{equation}

\begin{figure}[ht]
    \centering
    \includegraphics[width=6.8cm]{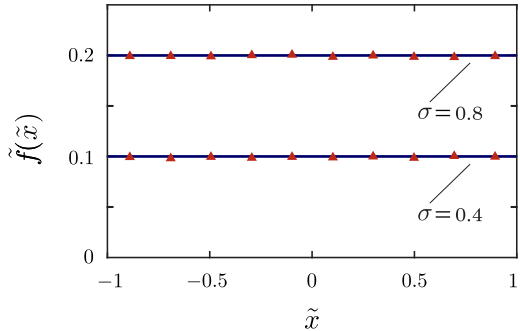}
    \caption{Reduced non-normalized probability
    density $\tilde{f}(\tilde{x})$ as a function of
    the reduced variable $\tilde{x} = x/l$ for
    $\sigma = 0.4$ and $\sigma = 0.8$. The solid
    horizontal lines represent the theoretical
    result \textcolor{blue} {(\ref{f1,a1})}
    for $\tilde{f}$, and triangle symbols represent
    the results of numerical simulations of \textcolor{blue}
    {Eq.\ (\ref{eq_X})}. The theoretical values
    of the probability $a$ ($a=0.4$ for $\sigma =
    0.4$ and $a=0.3$ for $\sigma = 0.8$) are also
    in good agreement with the numerical ones
    ($a \approx a_{-} \approx a_{+}$).}
    \label{fig1}
\end{figure}

Thus, at $n=1$ the non-normalized probability density is uniform, i.e.,  $\tilde{f} (\tilde{x}) = \tilde{f}$ for all $|\tilde{x}| \leq 1$ (the only one branch exists in this case). According to \textcolor{blue} {(\ref{f1,a1})}, the probability density $\tilde{f}$ decreases and the probability $a$ increases as the ratio parameter $\sigma$ decreases. For small $\sigma$, these results can be understood by noting that the mean value of $|z_{i}|$, which we denote as $Z$, is inversely proportional to $\sigma$. Indeed, since  $Z = \int_{-\infty}^ {\infty} |z|q(z)\,dz = l/ \sigma$, the higher is $Z$ (i.e., the lower is $\sigma$), the higher is the probability $a$ and hence the lower is the probability density $\tilde{f}$. As illustrated in \textcolor{blue}{Fig.\ \ref{fig1}}, the above theoretical results are in complete agreement with those obtained by solving \textcolor{blue}{Eq.\ (\ref{eq_X})} numerically.

In order to derive these and other numerical results, we proceed as follows (see also Ref.\ \textcolor{blue} {\cite{DeBy_PhysA2018}}). First, considering $\tau$ as the time step and assuming that $\tau = 10^{-3}$, $l=1$, $\lambda=1$ and $X_{0}=0$ (here, the model parameters are chosen to be dimensionless), from \textcolor{blue}{Eq.\ (\ref{eq_X})} we find $X_{M\tau}$ for $N = 10^{6}$ simulation runs. Because we are concerned with the stationary state, the number of steps is taken to be large enough: $M=10^{4}$. Then, the interval $(-1,1)$ is divided into $K=50$ subintervals of width $\delta = 2/K$, and the reduced non-normalized probability density is defined as $\tilde{f} (\overline{x}_{m}) = N_{m}/\delta N$ , where $\overline{x}_{m}$ is the middle position of the $m$-th subinterval, $m = \overline{1,K}$, and $N_{m}$ is the number of runs for which $X_{M\tau}$ belongs to the $m$-th subinterval. Finally, the probability $a$ is defined as $a = (a_{-} + a_{+})/2$, where $a_{-} = N_{-}/N$, $a_{+} = N_{+}/N$, and $N_{-}$ and $N_{+}$ are the number of runs for which $X_{M\tau} = -l$ and $X_{M\tau} = l$, respectively.

\subsection{Solution at \texorpdfstring{$\sigma \in (1,2)$}{Lg}}

If $n=2$, then $\sigma \in (1,2)$ and so $c \in (l, 2l)$. Therefore, in this case the non-normalized probability density $f(x)$ must satisfy \textcolor{blue}{Eqs.\ (\ref{f_eq2})}. Assuming that $x \in [-l, l-c)$ and taking into account that $\int_{-l}^{0} f(x')\,dx' = (1-2a)/2$, we can rewrite \textcolor{blue}{Eq.\ (\ref{f_eq2a})} in the form
\begin{equation}
    f(y) = \frac{1}{4c} + \frac{1}{2c}
    \int_{0}^{c-l}f(x')\,dx' + \frac{1}{2c}
    \int_{c-l}^{y+c}f(x')\,dx'.
    \label{f_eq2a1}
\end{equation}
Here, for convenience of future calculations, we temporarily replaced the variable $x$ by $y$. By differentiating \textcolor{blue}{Eq.\ (\ref{f_eq2a1})} with respect to $y$, we get the equation
\begin{equation}
    \frac{d}{dy}f(y) = \frac{1}
    {2c}f(y+c),
    \label{f_eq2a2}
\end{equation}
which belongs to a class of differential difference equations (see, e.g., Ref.\ \textcolor{blue} {\cite{BeCo1963}}).

If $x \in (l-c, c-l)$, then, using \textcolor{blue} {(\ref{f_eq2b})} and condition \textcolor{blue} {(\ref{def_a})}, we immediately find
\begin{equation}
    f(x) = \frac{1}{2c}.
    \label{f2b}
\end{equation}
With this result, \textcolor{blue}{Eq.\ (\ref{f_eq2a1})} is reduced to
\begin{equation}
    f(y) = \frac{1}{2c} - \frac{l}
    {4c^{2}} + \frac{1}{2c}
    \int_{c-l}^{y+c}f(x')\,dx'.
    \label{f_eq2a3}
\end{equation}

Finally, if $x \in (c-l, l]$, then it is reasonable to divide the interval of integration in \textcolor{blue}{Eq.\ (\ref{f_eq2c})} by three subintervals $(x-c, l-c)$, $(l-c, 0)$ and $(0, l]$. This, together with the above results \textcolor{blue}{(\ref{def_a})}, \textcolor{blue}{(\ref{f2b})} and condition $x=y+c$, permits us to represent \textcolor{blue}{Eq.\ (\ref{f_eq2c})} in the form
\begin{equation}
    f(y+c) = \frac{1}{2c} - \frac{l}
    {4c^{2}} + \frac{1}{2c}
    \int_{y}^{l-c}f(x')\,dx',
    \label{f_eq2c1}
\end{equation}
which, after differentiating with respect to $y$, yields the following differential difference equation:
\begin{equation}
    \frac{d}{dy}f(y + c) =
    - \frac{1}{2c}f(y).
    \label{f_eq2c2}
\end{equation}

A set of differential difference equations \textcolor{blue} {(\ref{f_eq2a2})} and \textcolor{blue} {(\ref{f_eq2c2})} determines the non-normalized probability density on the intervals $[-l, l-c)$ and $(c-l, l]$. Its remarkable feature is that it can be reduced (by a single differentiation of these equations with respect to $y$) to a set of independent ordinary differential equations
\begin{subequations}\label{f_eq2ac}
\begin{gather}
    \frac{d^{2}}{dy^{2}}f(y) +
    \frac{1}{4c^{2}}f(y) =0,
    \label{f_eq2a4}\\
    \frac{d^{2}}{dy^{2}}f(y + c) +
    \frac{1}{4c^{2}}f(y + c) =0.
    \label{f_eq2c3}
\end{gather}
\end{subequations}
Since $-y \in (c-l, l]$ and $f(-y) = f(y)$, \textcolor{blue}{Eq.\ (\ref{f_eq2c3})} is equivalent to \textcolor{blue}{Eq.\ (\ref{f_eq2a4})}. Therefore, returning to the variable $x$, from the equation
\begin{equation}
    \frac{d^{2}}{dx^{2}}f(x) +
    \frac{1}{4c^{2}}f(x) =0
    \label{f_eq2gen}
\end{equation}
we find the function $f(x)$ at $x \in [-l, l-c)$,
\begin{equation}
    f(x) = \alpha \cos{\frac{x}{2c}}
    + \beta \sin{\frac{x}{2c}}
    \label{f2a}
\end{equation}
($\alpha$ and $\beta$ are parameters to be determined), and at $x \in (c-l, l]$,
\begin{equation}
    f(x) = \alpha' \cos{\frac{x}{2c}}
    + \beta' \sin{\frac{x}{2c}}.
    \label{f2c}
\end{equation}
Taking also into account that $f(-x) = f(x)$, one can make sure that $\alpha' = \alpha$ and $\beta' = - \beta$. Thus, collecting the above results, for  the non-normalized probability density $f(x)$ we obtain a general representation
\begin{equation}
    f(x) =
    \left\{\! \begin{array}{ll}
    \alpha \cos{(x/2c)}
    + \beta \sin{(x/2c)},
    & x \in [-l, l-c),
    \\
    1/2c, & x \in (l-c, c-l),
    \\
    \alpha \cos{(x/2c)}
    - \beta \sin{(x/2c)},
    & x \in (c-l, l].
    \end{array}
    \right.
    \label{f2}
\end{equation}

To find the parameters $\alpha$ and $\beta$, we use \textcolor{blue}{Eq.\ (\ref{f_eq2a3})} with $y = x \in [-l, l-c)$. Substituting \textcolor{blue}{(\ref{f2})} into \textcolor{blue}{Eq.\ (\ref{f_eq2a3})}, we arrive to the equation
\begin{gather}
    \left[ \alpha \left(1 -\sin{\frac{1}{2}}
    \right) - \beta \cos{\frac{1}{2}} \right]
    \cos{\frac{x}{2c}} + \left[ \beta \left(1 +
    \sin{\frac{1}{2}}\right) - \alpha \cos{
    \frac{1}{2}} \right] \sin{\frac{x}{2c}}
    \nonumber \\[4pt]
    +\, \alpha \sin{\frac{c-l}{2c}} + \beta
    \cos{\frac{c-l}{2c}} - \frac{1}{2c} +
    \frac{l}{4c^{2}} = 0.
    \label{eq_ab1}
\end{gather}
It holds for all $x$ only if three conditions
\begin{subequations}
\begin{gather}
    \alpha \left(1 -\sin{\frac{1}{2}}
    \right) - \beta \cos{\frac{1}{2}} = 0,
    \quad
    \beta \left(1 + \sin{\frac{1}{2}}\right)
    - \alpha \cos{\frac{1}{2}} = 0,
    \label{eq_ab2}
    \\[4pt]
    \alpha \sin{\frac{c-l}{2c}} + \beta
    \cos{\frac{c-l}{2c}} - \frac{1}{2c} +
    \frac{l}{4c^{2}} = 0
    \label{eq_ab2'}
\end{gather}
\end{subequations}
are simultaneously satisfied. Since conditions in \textcolor{blue} {(\ref{eq_ab2})} are equivalent (this can be verified directly), we can consider one of them (e.g., the first one) and condition \textcolor{blue} {(\ref{eq_ab2'})} as a set of linear equations for $\alpha$ and $\beta$. The straightforward solution of these equations leads to
\begin{equation}
    \alpha = \frac{1}{l}\frac{\sigma(1- \sigma/4)\cos{[(\pi-1)/4]}}
    {4\cos{[(\sigma + \pi - 1)/4]}},
    \quad
    \beta = \frac{1}{l}\frac{\sigma(1-
    \sigma/4)\sin{[(\pi-1)/4]}}
    {4\cos{[(\sigma + \pi - 1)/4]}}.
    \label{alpha,beta}
\end{equation}
Formulas \textcolor{blue} {(\ref{f2})} and \textcolor{blue} {(\ref{alpha,beta})} completely determine the non-normalized probability density function $f(x)$ in the case when $\sigma \in (1,2)$. Since $f(x)$ is expressed in terms of trigonometric functions, integral in \textcolor{blue} {(\ref{f2})} can be calculated analytically, yielding
\begin{equation}
    a = \frac{\sigma}{4} - \sqrt{2}\,
    \frac{(1-\sigma/4) \sin{[(\sigma-1)
    /4]}}{\cos{[(\sigma + \pi - 1)/4]}}.
    \label{a2}
\end{equation}

For convenience of analysis, we rewrite the non-normalized probability density \textcolor{blue} {(\ref{f2})} in the reduced form
\begin{equation}
    \tilde{f}(\tilde{x}) =
    \left\{\! \begin{array}{ll}
    \alpha l\cos{(\sigma \tilde{x}/4)}
    + \beta l\sin{(\sigma \tilde{x}/4)},
    & x \in [-1, -\tilde{x}_{1}),
    \\
    \sigma/4, & x \in (-\tilde{x}_{1},
    \tilde{x}_{1}),
    \\
    \alpha l\cos{(\sigma \tilde{x}/4)}
    - \beta l\sin{(\sigma \tilde{x}/4)},
    & x \in (\tilde{x}_{1}, 1],
    \end{array}
    \right.
    \label{red_f2}
\end{equation}
where $\tilde{x}_{1} = |2/\sigma - 1|$ (this definition of $\tilde{x}_{1}$ will be used for $\sigma \in (2,3)$ as well). The properties of this probability density are surprising and unexpected. Indeed, in contrast to the previous case, in this case the function $\tilde{f} (\tilde{x})$ has three branches and it is discontinuous at $\tilde{x} = \pm \tilde{x}_{1}$. We emphasize that this qualitative change of the behavior of $\tilde{f} (\tilde{x})$ occurs when the ratio parameter $\sigma$ exceeds the critical one $\sigma_{\mathrm{cr}} = 1$. Using \textcolor{blue} {(\ref{red_f2})} and \textcolor{blue} {(\ref{alpha,beta})}, it can be shown that
\begin{equation}
    \tilde{f}(\pm 1) = \frac{\sigma}{4}\left(
    1 - \frac{\sigma}{4} \right), \quad
    \tilde{f}(\pm \tilde{x}_{1} \pm 0) = \frac{
    \sigma}{4}\left(1 - \frac{\sigma}{4} \right)
    \tan{\frac{\sigma + \pi -1}{4}}
    \label{f2_lim}
\end{equation}
and $\tilde{f}(\pm 1) < \tilde{f}(\pm \tilde{x}_{1} \pm 0) < \sigma/4$. With increasing $\sigma$ from $1$ to $2$, the width of the intervals $[-1, -\tilde{x}_{1})$ and $(\tilde{x}_{1}, 1]$, where $\tilde{f}(\tilde{x})$ nonlinearly depends on $\tilde{x}$, increases from $0$ to $1$, and the width of the interval $(-\tilde{x}_{1}, \tilde{x}_{1})$, where $\tilde{f}(\tilde{x})$ does not depend on $\tilde{x}$, decreases from $2$ to $0$.

\begin{figure}[ht]
    \centering
    \includegraphics[width=\columnwidth]{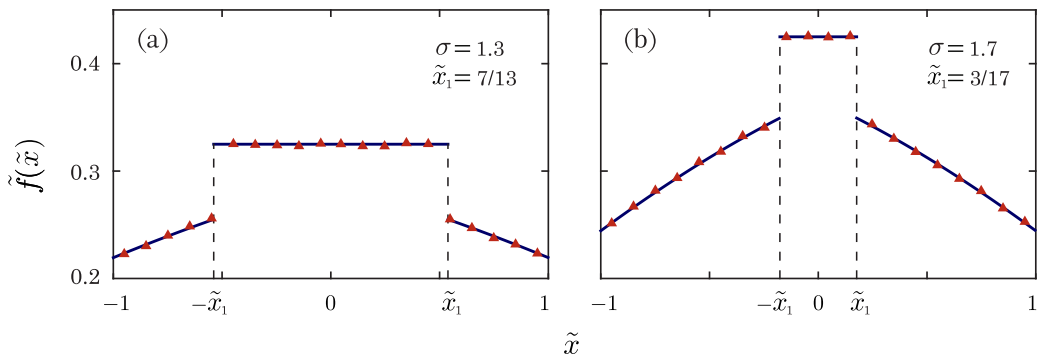}
    \caption{Reduced non-normalized probability
    density $\tilde{f}(\tilde{x})$ as a function of
    the reduced variable $\tilde{x} = x/l$ for
    $\sigma = 1.3$ (a) and $\sigma = 1.7$ (b). The
    solid lines show theoretical dependencies
    obtained from \textcolor{blue}{ (\ref{red_f2})}
    and \textcolor{blue} {(\ref{alpha,beta})},
    and the triangle symbols indicate the results
    obtained by numerical simulations of
    \textcolor{blue}{Eq.\ (\ref{eq_X})}.}
    \label{fig2}
\end{figure}

For the sake of illustration, in \textcolor{blue}{Fig.\ \ref{fig2}} we show the behavior of the reduced non-normalized probability density \textcolor{blue}{(\ref{red_f2})} for two values of the ratio parameter $\sigma$ (solid lines). In order to verify these theoretical results, we performed numerical simulations of \textcolor{blue}{Eq.\ (\ref{eq_X})}, paying a special attention to the vicinities of the points of discontinuity $\pm \tilde{x}_{1}$. As seen from this figure, the numerical results (denoted by triangle symbols) are fully consistent with the theoretical ones.

\subsection{Solution at \texorpdfstring{$\sigma \in (2,3)$}{Lg}}

To determine the non-normalized probability density $f(x)$ at $n=3$, i.e., when $\sigma \in (2,3)$ or, equivalently, when $c \in (2l/3, l)$, we should use \textcolor{blue}{Eqs.\ (\ref{f_eq3})}. Since in this case the chain of inequalities $-l < l-2c < c-l < l-c < 2c-l < l$ holds, it is reasonable to divide the interval $[-l, l]$ into five subintervals $[-l, l-2c)$, $(l-2c, c-l)$, $(c-l, l-c)$, $(l-c, 2c-l)$ and $(2c-l, l]$. Then, using formula \textcolor{blue}{(\ref{def_a})}, from \textcolor{blue}{Eq.\ (\ref{f_eq3a})} one can derive the equations
\begin{equation}
    f(y_{1}) = \frac{1}{4c} + \frac{1}
    {2c} \int_{0}^{y_{1}+c}f(x')\,dx'
    \label{f_eq3a1}
\end{equation}
and
\begin{equation}
    \frac{d}{dy_{1}}f(y_{1}) = \frac{1}{2c}
    f(y_{1}+c),
    \label{f_eq3a2}
\end{equation}
if $y_{1} \in [-l, l-2c)$, and the equations
\begin{equation}
    f(y_{2}) = \frac{1}{4c} + \frac{1}
    {2c} \int_{0}^{l-c}f(x')\,dx' +
    \frac{1}{2c} \int_{l-c}^{y_{2}+c}f(x')
    \,dx'
    \label{f_eq3a3}
\end{equation}
and
\begin{equation}
    \frac{d}{dy_{2}}f(y_{2}) = \frac{1}{2c}
    f(y_{2}+c),
    \label{f_eq3a4}
\end{equation}
if $y_{2} \in (l-2c, c-l)$ (we temporary use the variables $y_{1}$ and $y_{2}$ instead of the variable $x$).

Similarly, \textcolor{blue}{Eq.\ (\ref{f_eq3b})} at $x = y_{1} +c \in (c-l, l-c)$ yields the equation
\begin{equation}
    f(y_{1}+c) = \frac{1}{2c}(1-2a) - \frac{1}
    {2c} \int_{-l}^{y_{1}}f(x')\,dx' - \frac{1}
    {2c} \int_{y_{1} + 2c}^{l}f(x')\,dx',
    \label{f_eq3b1}
\end{equation}
from which one immediately obtains
\begin{equation}
    \frac{d}{dy_{1}}f(y_{1} + c) = -\frac{1}{2c}
    f(y_{1}) + \frac{1}{2c} f(y_{1} + 2c).
    \label{f_eq3b2}
\end{equation}

Finally, from \textcolor{blue}{Eq.\ (\ref{f_eq3c})} we find the equations
\begin{equation}
    f(y_{2}+c) = \frac{1}{4c} + \frac{1}
    {2c} \int_{c-l}^{0}f(x')\,dx' +
    \frac{1}{2c}\int_{y_{2}}^{c-l}f(x')\,dx'
    \label{f_eq3c1}
\end{equation}
and
\begin{equation}
    \frac{d}{dy_{2}}f(y_{2} +c) = -\frac{1}{2c}
    f(y_{2}),
    \label{f_eq3c2}
\end{equation}
if $x = y_{2} + c \in (l-c, 2c-l)$, and the equations
\begin{equation}
    f(y_{1} + 2c) = \frac{1}{4c} + \frac{1}
    {2c} \int_{y_{1} + c}^{0}f(x')\,dx'
    \label{f_eq3c3}
\end{equation}
and
\begin{equation}
    \frac{d}{dy_{1}}f(y_{1} +2c) = -\frac{1}{2c}
    f(y_{1}+c),
    \label{f_eq3c4}
\end{equation}
if $x = y_{1} + 2c \in (2c-l, l]$.

Let us first consider two sets of the above differential difference equations, namely, a set of \textcolor{blue}{Eqs.\ (\ref{f_eq3a2})}, \textcolor{blue} {(\ref{f_eq3b2})} and \textcolor{blue} {(\ref{f_eq3c4})}, and a set of \textcolor{blue}{Eqs.\ (\ref{f_eq3a4})} and \textcolor{blue} {(\ref{f_eq3c2})}. Remarkably, each of these sets can also be reduced to a set of independent ordinary differential equations that are easily solved. In particular, by differentiating \textcolor{blue}{Eq.\ (\ref{f_eq3b2})} with respect to $y_{1}$ and using \textcolor{blue}{Eqs.\ (\ref{f_eq3a2})} and \textcolor{blue} {(\ref{f_eq3c4})}, we get
\begin{equation}
    \frac{d^{2}}{dy_{1}^{2}}f(y_{1} + c)
    + \frac{1}{2c^{2}} f(y_{1} + c) =0.
    \label{eq_f1}
\end{equation}
Returning to the variable $x = y_{1} + c$, the symmetric solution of this equation can be represented as
\begin{equation}
    f(x) = \mu \cos{\frac{x}{\sqrt{2}c}},
    \label{f3a}
\end{equation}
where $x \in (c-l, l-c)$ and $\mu$ is a parameter to be determined. Then, substituting $f(y_{1} + c)$ from \textcolor{blue}{Eq.\ (\ref{f_eq3a2})} into \textcolor{blue}{Eq.\ (\ref{eq_f1})} and returning to the variable $x$, one obtains the equation
\begin{equation}
    \frac{d^{3}}{dx^{3}}f(x)
    + \frac{1}{2c^{2}} \frac{d}
    {dx} f(x) =0,
    \label{eq_f2}
\end{equation}
which holds for both $x \in [-l, l-2c)$ and $x \in (2c-l, l]$. Using the symmetry condition $f(-x) = f(x)$, the solution of this equation can be written in the form
\begin{equation}
    f(x) = \left\{\! \begin{array}{ll}
    \eta \cos{(x/\!\sqrt{2}c)}
    + \kappa \sin{(x/\!\sqrt{2}c)}
    + \gamma, & x \in [-l, l-2c),
    \\
    \eta \cos{(x/\!\sqrt{2}c)}
    - \kappa \sin{(x/\!\sqrt{2}c)}
    + \gamma, & x \in (2c-l, l].
    \end{array}
    \right.
    \label{f3b}
\end{equation}

Similarly, it can be shown that the set of \textcolor{blue}{Eqs.\ (\ref{f_eq3a4})} and \textcolor{blue} {(\ref{f_eq3c2})} is reduced to \textcolor{blue}{Eq.\ (\ref{f_eq2gen})}, which holds on intervals $(l-2c, c-l)$ and $(l-c, 2c-l)$. The solution of this equation, satisfying the condition $f(-x) = f(x)$, is given by
\begin{equation}
    f(x) = \left\{\! \begin{array}{ll}
    \nu \cos{(x/2c)} + \chi \sin{(x/2c)},
    & x \in (l-2c, c-l),
    \\
    \nu \cos{(x/2c)} - \chi \sin{(x/2c)},
    & x \in (l-c, 2c-l).
    \end{array}
    \right.
    \label{f3c}
\end{equation}

To determine the unknown parameters in \textcolor{blue} {(\ref{f3a})}, \textcolor{blue} {(\ref{f3b})} and \textcolor{blue} {(\ref{f3c})}, we use \textcolor{blue}{Eqs.\ (\ref{f_eq3a1})}, \textcolor{blue} {(\ref{f_eq3a3})} and \textcolor{blue} {(\ref{f_eq3b1})}. Substituting $f(x)$ from \textcolor{blue} {(\ref{f3a})}, \textcolor{blue} {(\ref{f3b})} and \textcolor{blue} {(\ref{f3c})} into these equations and omitting technical details, we obtain the following representation for the non-normalized probability density:
\begin{equation}
    f(x) =
    \left\{\! \begin{array}{ll}
    (\mu/\!\sqrt{2})\sin{[(c+x)/\!\sqrt{2}c]}
    + 1/4c, & x \in [-l, l-2c),
    \\
    \nu \left( \cos{(x/2c)} + \sin
    {[(c+x)/2c]}\right),
    & x \in (l-2c, c-l),
    \\
    \mu\cos{(x/\!\sqrt{2}c)},
    & x \in (c-l, l-c),
    \\
    \nu \left( \cos{(x/2c)} + \sin
    {[(c-x)/2c]}\right),
    & x \in (l-c, 2c-l),
    \\
    (\mu/\!\sqrt{2})\sin{[(c-x)/\!\sqrt{2}c]}
    + 1/4c, & x \in (2c-l, l],
    \end{array}
    \right.
    \label{f3}
\end{equation}
where
\begin{equation}
    \mu = \frac{\sigma}{8l}\frac{1}
    {\sin{[(\sigma/2 -1)/\!\sqrt{2}]}}
    \frac{\sigma/2 -3 + 2
    \cot{[(\sigma + \pi -3) /4]}}
    {\cot{[(\sigma/2 -1)/
    \!\sqrt{2}]} - \!\sqrt{2}
    \cot{[(\sigma + \pi -3)/4]}}
    \label{mu}
\end{equation}
and
\begin{equation}
    \nu = \frac{\sigma}{16l}\frac{\cos{(1/4)}
    - \sin{(1/4)}} {\cos{(1/2)}\sin{[(\sigma +
    \pi -3) /4]}} \frac{\sigma/2 -3 +
    \!\sqrt{2} \cot{[(\sigma/2 -1) /\!\sqrt{2}]}}
    {\cot{[(\sigma/2 -1)/\!\sqrt{2}]} - \!\sqrt{2}
    \cot{[(\sigma + \pi -3)/4]}}.
    \label{nu}
\end{equation}
Finally, by direct integration of $f(x)$, from \textcolor{blue} {(\ref{def_a})} one gets
\begin{equation}
    a = \frac{3}{2} - \frac{\sigma}{4}
    - \frac{\sqrt{2}}{4} \frac{\sigma/2 -3 +\sqrt{2}
    \cot{[(\sigma/2 -1) /\!\sqrt{2}]}}{\cot{[
    (\sigma/2 -1)/\!\sqrt{2}]} - \!\sqrt{2} \cot{
    [(\sigma + \pi -3)/4]}} \cot{\frac{\sigma +
    \pi -3}{4}}.
    \label{a3}
\end{equation}

\begin{figure}[ht]
    \centering
    \includegraphics[width=\columnwidth]{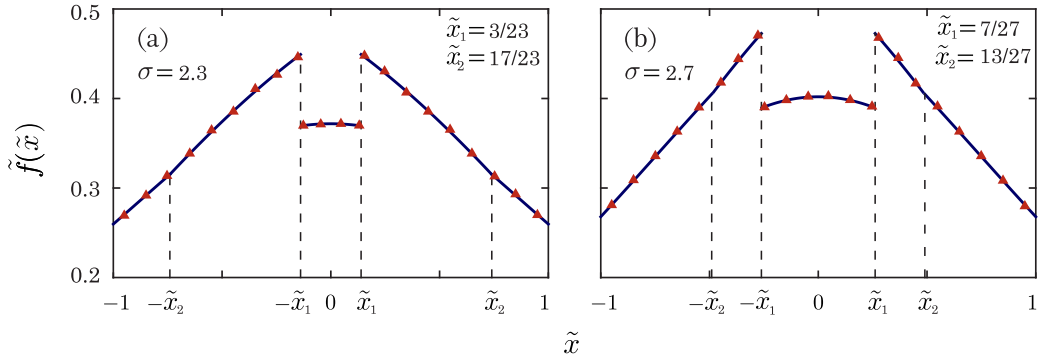}
    \caption{Reduced non-normalized probability
    density $\tilde{f}(\tilde{x})$ as a function of
    the reduced variable $\tilde{x} = x/l$ for
    $\sigma = 2.3$ (a) and $\sigma = 2.7$ (b). The
    solid lines represent the theoretical results
    obtained using \textcolor{blue}{(\ref{red_f3})},
    \textcolor{blue}{(\ref{mu})} and \textcolor{blue}
    {(\ref{nu})}, and the triangle symbols show the
    numerical results obtained by numerical
    simulations of \textcolor{blue}{Eq.\ (\ref{eq_X})}.}
    \label{fig3}
\end{figure}

In the reduced form, the non-normalized probability density \textcolor{blue} {(\ref{f3})} is rewritten as
\begin{equation}
    \tilde{f}(\tilde{x}) =
    \left\{\! \begin{array}{ll}
    (\mu l/\!\sqrt{2})\sin{[(c+x)/\!\sqrt{2}c]}
    + \sigma/8, & x \in [-1, -\tilde{x}_{2}),
    \\
    \nu l\left( \cos{(x/2c)} + \sin
    {[(c+x)/2c]}\right),
    & x \in (-\tilde{x}_{2}, -\tilde{x}_{1}),
    \\
    \mu l\cos{(x/\!\sqrt{2}c)},
    & x \in (-\tilde{x}_{1}, \tilde{x}_{1}),
    \\
    \nu l\left( \cos{(x/2c)} + \sin
    {[(c-x)/2c]}\right),
    & x \in (\tilde{x}_{1}, \tilde{x}_{2}),
    \\
    (\mu l/\!\sqrt{2})\sin{[(c-x)/\!\sqrt{2}c]}
    + \sigma/8, & x \in (\tilde{x}_{2}, 1],
    \end{array}
    \right.
    \label{red_f3}
\end{equation}
where $\tilde{x}_{2} = |4/\sigma - 1|$ and $\tilde{x}_{1} < \tilde{x}_{2} < 1$. In accordance with the general rule formulated at the end of \textcolor{blue}{Section \ref{BasEq}}, in this case the function $\tilde{f}( \tilde{x})$ has five branches separated from each other by four points $\pm \tilde{x}_{1}$ ($\tilde{f}( \tilde{x})$ is discontinuous at $\tilde{x} = \pm \tilde{x}_{1}$) and $\pm \tilde{x}_{2}$ ($\tilde{f}( \tilde{x})$ is continuous at $\tilde{x} = \pm \tilde{x}_{2}$). The intervals $[-1, -\tilde{x}_{2})$, $(-\tilde{x}_{1}, \tilde{x}_{1})$ and $(\tilde{x}_{2}, 1]$ have the same width $2 - 4/\sigma$, which increases from $0$ to $2/3$ as the ratio parameter $\sigma$ grows from $2$ to $3$. In contrast, the width $6/\sigma - 2$ of the intervals $(-\tilde{x}_{2}, -\tilde{x}_{1})$ and $(\tilde{x}_{1}, \tilde{x}_{2})$ decreases from $1$ to $0$. As in the previous cases, the theoretical results obtained for $\sigma \in (2,3)$ are confirmed by numerical simulations, see \textcolor{blue}{Fig.\ \ref{fig3}}.

\subsection{Solution at \texorpdfstring{$\sigma \to \infty$}{Lg}}

The above results indicate that, because the number of branches of the non-normalized PDF $f(x)$ grows, its local behavior becomes more and more complex with increasing parameter $\sigma$ (we recall, $\sigma$ is the ratio of the domain size $2l$ of the bounded process $X_{t}$ to the half-width $c$ of uniform distribution of jump magnitudes $z_{i}$). For this reason, we were not able to solve \textcolor{blue}{Eqs.\ (\ref{f_eq3})} analytically for arbitrary large values of $\sigma$ (we solved it for $n=4$ as well, but the results are too cumbersome to present here). However, the function $f(x)$ in the limit $\sigma \to \infty$ approaches a constant, which can be determined as follows. First, using \textcolor{blue} {(\ref{q2})}, we find
\begin{equation}
    \int_{-l}^{l} q(x-x')\,dx' =
    \frac{1}{2c} \times
    \left\{\! \begin{array}{ll}
    l+c+x, & x \in [-l, c-l],
    \\
    2c, & x \in [c-l, l-c],
    \\
    l+c-x, & x \in [l-c, l].
    \end{array}
    \right.
    \label{Int_q}
\end{equation}
Then, assuming that $f(x) = h = \mathrm{const}$ and substituting expressions \textcolor{blue} {(\ref{q2})} and \textcolor{blue} {(\ref{Int_q})} into \textcolor{blue}{Eq.\ (\ref{f_eq})}, one can make sure that at $x \in (-l+c, l-c)$ this equation is satisfied identically, and at $x \in [-l, -l+c]$ it reduces to
\begin{equation}
    h = \frac{a}{2c} + h\, \frac{l+c+x}{2c}.
    \label{f_eq4}
\end{equation}
(Note, at $x \in [l-c, l]$ \textcolor{blue}{Eq.\ (\ref{f_eq})} reduces to \textcolor{blue}{Eq.\ (\ref{f_eq4})} with $x$ replaced by $-x$.)

As it follows from \textcolor{blue}{Eq.\ (\ref{f_eq4})}, our assumption that $f(x)$ does not depend on $x$ is, strictly speaking, incorrect. Nevertheless, if $c \ll l$ (i.e., $\sigma \gg 1$), it can be used as a first approximation. Indeed, taking into account that, according to \textcolor{blue} {(\ref{def_a})}, $a = 1/2 - hl$, from \textcolor{blue}{Eq.\ (\ref{f_eq4})} one obtains
\begin{equation}
    h = \frac{1}{2l} \frac{1}{1 +
    (c-l-x)/l}.
    \label{f_4}
\end{equation}
Since the values of $c-l-x$ for $x \in [-l, -l+c]$ belong to the interval $[0,c]$ and the condition $c \ll l$ holds, we get $h = 1/2l$ and $a=0$ as $\sigma \to \infty$. Hence, in this limit the reduced PDF \textcolor{blue} {(\ref{red_Pst})} takes the form
\begin{equation}
    \tilde{P}_{\mathrm{st}}(\tilde{x})\,
    |_{\sigma \to \infty}
    = \frac{1}{2} \quad (|\tilde{x}| \leq 1).
    \label{Pst_lim}
\end{equation}
Our numerical simulations show that (if $\sigma \gtrsim 50$) this result is reproduced with an accuracy of a few percent or better [to estimate the accuracy analytically, one can use formula \textcolor{blue} {(\ref{h})}]. It should be noted in this regard that with increasing $\sigma$ the number of time steps $M$, which is necessary to reach the stationary state, increases as well. We also stress that the same result \textcolor{blue} {(\ref{Pst_lim})} holds for the bounded process $X_{t}$ driven by Gaussian white noise \textcolor{blue} {\cite{Gard2009}}.

\subsection{Extreme values probability}

\begin{figure}[ht]
    \centering
    \includegraphics[width=6.8cm]{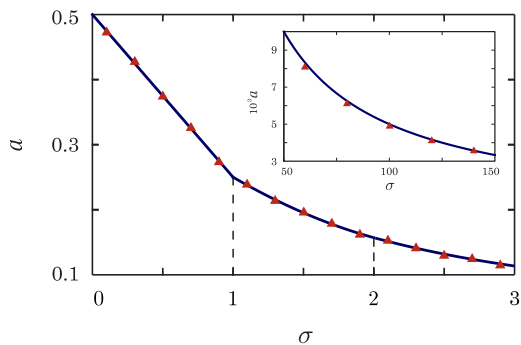}
    \caption{Probability $a$ of the extremal
    values of the bounded process $X_{t}$ as
    a function of the ratio parameter $\sigma$.
    The solid lines represent the theoretical
    results \textcolor{blue}{(\ref{f1,a1})},
    \textcolor{blue} {(\ref{a2})} and \textcolor{blue}
    {(\ref{a3})} for $\sigma \in (0,1)$, $\sigma
    \in (1,2)$ and $\sigma \in (2,3)$,
    respectively. The results obtained via
    numerical simulations of \textcolor{blue}{Eq.\
    (\ref{eq_X})} are marked by triangle symbols.
    Inset: probability $a$ vs.\ $\sigma$ for large
    values of $\sigma$. The solid line represents
    the asymptotic formula $a=1/2\sigma$.}
    \label{fig4}
\end{figure}

The probability $a$ that in the stationary state $X_{t} = -l$ (or $X_{t} = l$) is determined by the formulas \textcolor{blue} {(\ref{f1,a1})}, \textcolor{blue} {(\ref{a2})} and \textcolor{blue} {(\ref{a3})} for $\sigma \in (0,1)$, $\sigma \in (1,2)$ and $\sigma \in (2,3)$, respectively. Using these formulas, it can be directly shown that $a|_{\sigma=1-0} = a|_{\sigma=1+0}$ and $a|_{\sigma=2-0} = a|_{\sigma=2+0}$, i.e., $a$ is a continuous function of $\sigma$ at the critical points $\sigma_{\mathrm{ cr}} = 1$ and $\sigma_{\mathrm{ cr}} = 2$, and $a$ monotonically decreases as the ratio parameter $\sigma$ increases from $0$ to $3$. As \textcolor{blue}{Fig.\ \ref{fig4}} illustrates, our theoretical results \textcolor{blue} {(\ref{f1,a1})}, \textcolor{blue} {(\ref{a2})} and \textcolor{blue} {(\ref{a3})} are in excellent agreement with the simulation data.

Although we have no explicit expressions for the probability $a$ at $\sigma > 3$, it can be easily seen that $a$ as a function of $\sigma$ is continuous at the critical points $\sigma_{\mathrm{ cr}} = \overline{3,\infty}$ as well. Indeed, according to the properties of $f(x)$ formulated in \textcolor{blue}{Section \ref{BasEq}}, the function $f(x)$ at $\sigma = \sigma_{\mathrm{ cr}} + 0$ acquires new branches (compared to $\sigma = \sigma_{\mathrm{ cr}} - 0$), which are located at separate points. Since these points do not contribute to the integral in \textcolor{blue} {(\ref{def_a})}, one may conclude that $a|_{\sigma = \sigma_{\mathrm{ cr}} - 0} = a|_{\sigma = \sigma_{ \mathrm{ cr}} + 0}$, i.e., the probability $a$ as a function of the ratio parameter $\sigma$ is continuous at all critical points $\sigma = \sigma_{\mathrm{ cr}}$.

Using results of the previous section, we can also estimate the dependence of $a$ on $\sigma$ for $\sigma \gg 1$. To this end, we first note that, according to our assumption $f(x) = h = \mathrm{const}$, the condition $\int_{-l}^{l} f(x)\,dx = 2lh$ must hold. On the other hand, from the above results it follows that
\begin{equation}
    \int_{-l}^{l} f(x)\,dx = 2(l-c)h + \frac{1}{2}
    \int_{-l}^{-l+c} \frac{dx}{c-x} + \frac{1}{2}
    \int_{l-c}^{l} \frac{dx}{c+x} .
    \label{Int}
\end{equation}
Performing integration and equating the right-hand side of \textcolor{blue} {(\ref{Int})} to $2lh$, we obtain
\begin{equation}
    h = \frac{\sigma}{4l} \ln{\! \bigg( 1 +
    \frac{2}{\sigma} \bigg)}
    \label{h}
\end{equation}
and, since $a = 1/2 - lh$,
\begin{equation}
    a = \frac{1}{2} - \frac{\sigma}{4} \ln{\! \bigg( 1 +
    \frac{2}{\sigma} \bigg)}.
    \label{a4}
\end{equation}
Taking into account that the ratio parameter $\sigma$ is assumed to be large enough, from \textcolor{blue} {(\ref{a4})} one gets in the first nonvanishing approximation: $a = 1/2\sigma$ as $\sigma \to \infty$. Our numerical results confirm this theoretical prediction, see inset in \textcolor{blue}{Fig.\ \ref{fig4}} (note, to reach the stationary state at $\sigma \in (50,150)$, the number of steps $M$ was chosen to be $3\cdot 10^{8}$).

\section{Conclusions}
\label{Concl}

We have studied the statistical properties of a class of bounded jump processes governed by a special case of the difference Langevin equation driven by Poisson white noise, i.e., a random sequence of delta pulses. In contrast to the ordinary Langevin equation, this equation, due to the use of the saturation function, has only bounded solutions. We have derived the Kolmogorov-Feller equation for the normalized probability density function (PDF) of these processes and found its stationary solutions in the case of the uniform distribution of pulse sizes, which is assumed to be symmetric. It has been explicitly shown that the stationary PDF can be decomposed into two singular terms defining the probability of the process extreme values and a regular part representing the non-normalized PDF inside a bounded domain. Amazingly, the non-normalized PDF has proven to be a complex piecewise function with jump discontinuities.

One of the most remarkable findings is that the ratio of the width of the saturation function to the half-width of the uniform distribution of pulse sizes is the only parameter which controls all properties of the stationary PDF. In particular, the ratio parameter determines the number of branches of the non-normalized PDF and coordinates of points separating these branches. It has been also established that, with its increasing, two new branches are created every time the ratio parameter is equal to a natural number. Interestingly, although this enhances the local complexity of the stationary PDF, it approaches a constant in the limit of large values of the ratio parameter. All our theoretical predictions have been confirmed by numerical simulations of the difference Langevin equation.

To the best of our knowledge, the proposed Langevin model of bounded jump processes driven by Poisson white noise is the first one that allows to study the nontrivial statistical properties of these processes in great analytical detail.

\section*{Acknowledgment}

This work was partially supported by the Ministry of Education and Science of Ukraine under Grant No. 0119U100772.


\end{document}